\documentclass{article}
\usepackage{silence}
\usepackage[english]{babel}
\usepackage[utf8]{inputenc}
\usepackage{amsmath,amssymb}
\usepackage{enumitem}
\usepackage{lipsum}

\usepackage{setspace}

\usepackage{microtype,graphicx,subfigure}
\usepackage{booktabs} 

\usepackage{multirow}

\usepackage{tikz}
\usetikzlibrary{shapes.misc,patterns}
\usetikzlibrary{arrows.meta}
\usetikzlibrary{positioning,fit}
\usepackage{pgfplots}
\pgfplotsset{compat=1.9}

\usepackage{hyperref}

\usepackage[accepted]{icml2018}

\renewcommand{\vec}{\boldsymbol}

\newcommand{\T}{^\top}
\newcommand{\inv}{^{-1}}

\newcommand{\R}{\mathbb{R}}

\newcommand{\E}{\mathbb{E}}
\newcommand{\V}{\mathbb{V}}
\newcommand{\N}{\mathcal{N}}
\newcommand{\M}{\mathcal{M}}

\newcommand{\Dexp}{\mathcal{D}_{\mathrm{exp}}}
\newcommand{\Dsim}{\mathcal{D}_{\mathrm{sim}}}
\newcommand{\Dsimi}{\mathcal{D}_{\mathrm{sim},i}}

\newcommand{\GP}{\mathcal{GP}}

\newcommand{\thetamap}{\hat{\vec \theta}}
\newcommand{\mumarg}{\breve{\vec \mu}}
\newcommand{\mumarge}{\breve{\mu}_{(e)}}
\newcommand{\Sigmamarg}{\breve{\vec \Sigma}}
\newcommand{\td}[1]{\tilde{#1}}

\newcommand{\DBH}{D_{\mathrm{BH}}}
\newcommand{\DBF}{D_{\mathrm{BF}}}
\newcommand{\DAW}{D_{\mathrm{AW}}}

\newcommand{\Drandom}{\mathrm{Uni.}}

\newcommand{\diag}{\mathrm{diag}}
\newcommand{\tr}{\mathrm{tr}}

\DeclareMathOperator*{\argmax}{arg\,max}

\newlength{\plotwidth}
\newlength{\plotheight}
\newlength{\pkplotwidth}
\newlength{\pkplotheight}
\tikzset{
  base/.style = {rectangle, rounded corners, draw=black, fill=none, minimum width=10mm, minimum height=6mm,text centered},
}

\newcommand{\tablespace}{\\[1.25mm]}
\newcommand\Tstrut{\rule{0pt}{2.6ex}}         
\newcommand\tstrut{\rule{0pt}{2.0ex}}         
\newcommand\Bstrut{\rule[-0.9ex]{0pt}{0pt}}   

\usepackage{color,soul}
\newcommand{\new}[1]{\hl{#1}}
\newcommand{\couldremove}[1]{{\sethlcolor{red}\hl{#1}}}
\setstcolor{red}
\newcommand{\remove}[1]{\st{#1}}

\soulregister\cite7
\soulregister\citet7
\soulregister\citep7
\soulregister\ref7
\soulregister\pageref7
\soulregister\fig7
\soulregister\eq7
\soulregister\eqref7
\soulregister\sect7
\soulregister\ul7
\soulregister\remove7

\usepackage{calc,xcolor}

\renewcommand{\new}[1]{#1}
\renewcommand{\couldremove}[1]{#1}
\renewcommand{\remove}[1]{}

\icmltitlerunning{Design of Experiments for Model Discrimination Hybridising Analytical and Data-Driven Approaches}
\begin{document}
\twocolumn[
\icmltitle{Design of Experiments for Model Discrimination\\Hybridising Analytical and Data-Driven Approaches}

\begin{icmlauthorlist}
	\icmlauthor{Simon Olofsson}{ic}
	\icmlauthor{Marc Peter Deisenroth}{ic,pr}
	\icmlauthor{Ruth Misener}{ic}
\end{icmlauthorlist}
\icmlaffiliation{ic}{Dept. of Computing, Imperial College London, United Kingdom.}
\icmlaffiliation{pr}{PROWLER.io, United Kingdom}
\icmlcorrespondingauthor{Simon Olofsson}{simon.olofsson15@imperial.ac.uk}
\icmlcorrespondingauthor{Marc Peter Deisenroth}{m.deisenroth@imperial.ac.uk}
\icmlcorrespondingauthor{Ruth Misener}{r.misener@imperial.ac.uk}

\icmlkeywords{Machine Learning, ICML}
\vskip 0.3in
]
\printAffiliationsAndNotice{} 

\begin{abstract}
Healthcare companies must submit pharmaceutical drugs or medical devices to regulatory bodies before marketing new technology. Regulatory bodies frequently require transparent and interpretable computational modelling to justify a new healthcare technology, but researchers may have several competing models for a biological system and too little data to discriminate between the models.
In design of experiments for model discrimination, the goal is to design maximally informative physical experiments in order to discriminate between rival predictive models. Prior work has focused either on analytical approaches, which cannot manage all functions, or on data-driven approaches, which may have computational difficulties or lack interpretable marginal predictive distributions.
We develop a methodology introducing Gaussian process surrogates in lieu of the original mechanistic models.
We thereby extend existing design and model discrimination methods developed for analytical models to cases of non-analytical models in a computationally efficient manner.
\end{abstract}

\section{Introduction}
A patient's health and safety is an important concern, and healthcare is a heavily regulated industry. In the USA and the EU, private and public regulatory bodies exist on federal/union\remove{, state and local}\new{ and state} level\new{s} \citep{Field2008, Hervey2010}. Healthcare companies applying to market a new drug or medical device must submit extensive technical information to the regulatory bodies. In the USA and the EU, the Food \& Drug Administration (FDA) and European Medicines Agency (EMA) handle these applications, respectively. The FDA require that the technical information contains e.g.\ the chemical composition of the drug, how the drug affects the human body (pharmacodynamics), how the body affects the drug (pharmacokinetics), and methods for drug manufacturing, packaging and quality control \citep{FDA_CFR_21a_drugs}. Likewise, applying to market a new medical device requires submitting extensive technical information. A medical device can be any appliance, software or material used for non-pharmacological diagnosis, monitoring or treatment of a disease or condition \citep{EU_MD_regulations}. An important part of proving the efficacy and safety of a new drug or medical device is showing that its effects on the patient can be predicted and interpreted, e.g.\ via mathematical and computational models.

\begin{figure*}[!t]
	\begin{center}
    	\includegraphics[width=0.97\textwidth]{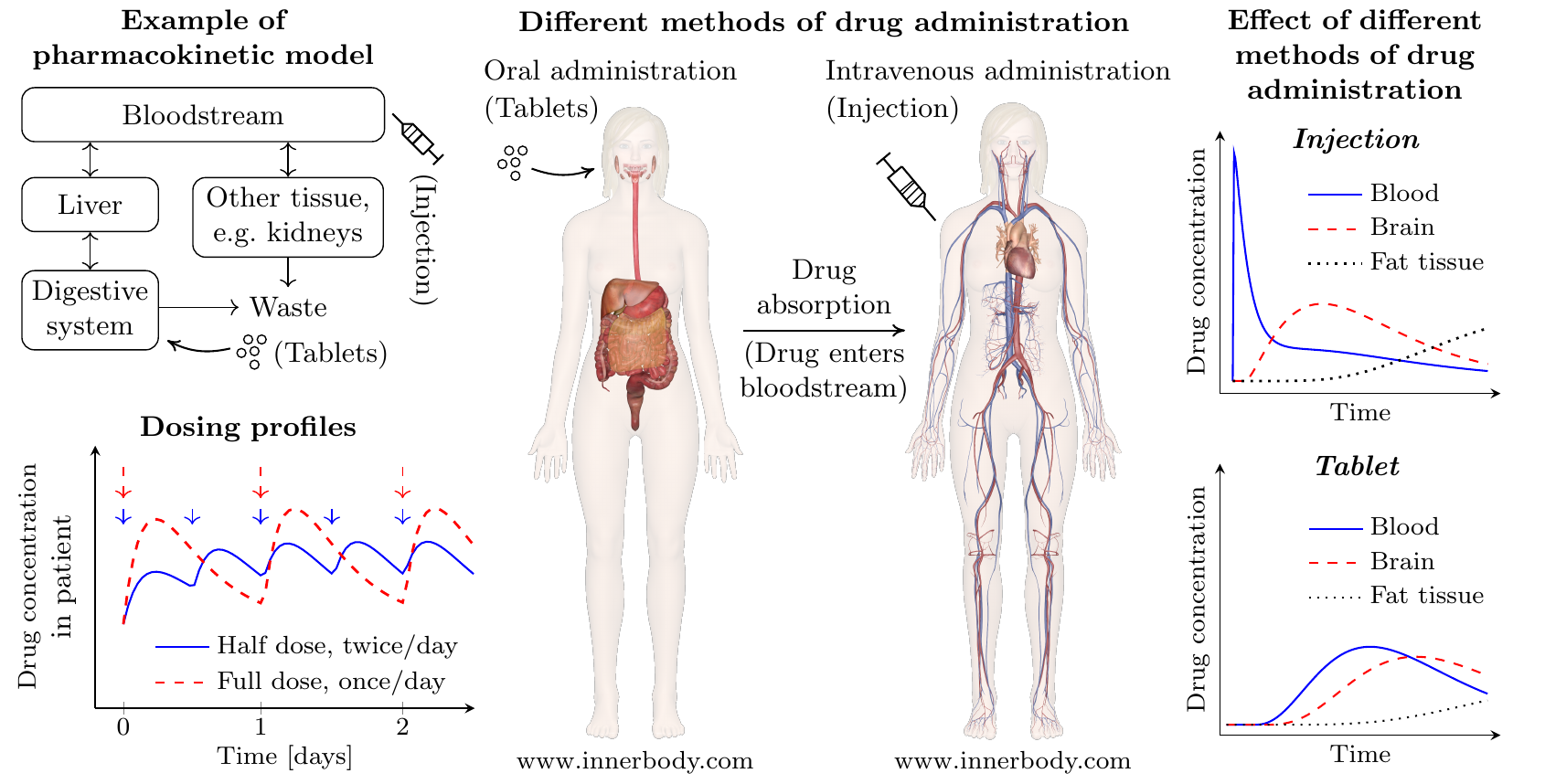}
		\vskip -0.05in
    	\caption{Different aspects of pharmacokinetic models. (Top left) \remove{example of a }simple pharmacokinetic model with four compartments: digestive system, liver, bloodstream and other tissue. The drug can enter as tablets into the digestive system or as an injection into the bloodstream, and leave the system by being excreted as waste. (Bottom left) \remove{example of }the effect on drug concentration in a patient from two different dosing profiles: \remove{giving the patient}\new{(i)} half the dose twice per day, or \new{(ii) }the full dose once per day. (Centre) two different methods of drug administration: orally in the form of tablets, or intravenously as an injection\remove{.}\new{, and} (Right) \remove{example of the effect of two different methods of drug administration, where}\new{their different effects:} an injection has a\remove{ much} quicker effect than \remove{administrating the drug through }tablets.}
    	\label{fig:pk}
	\end{center}
    \vspace{-3mm}
\end{figure*}

These transparency and interpretability requirements\new{, combined}\remove{, together} with limited amounts of available experimental data, make models learnt solely from observed \new{data\remove{ (e.g. neural networks)} unsuitable} for proving the efficacy and safety of a new drug or medical device to the regulatory bodies. Hence researchers use explicit parametric models. For drugs, the pharmacokinetics (how the human body affects the drug) is often modelled using systems of ordinary differential equations. These models elucidate how the drug is absorbed and distributed through the body (e.g. brain, kidneys and liver) under different dosing profiles and methods of drug administration (e.g.\ orally or intravenously) (see Figure~\ref{fig:pk}). The EMA\remove{ state in their} pharmacokinetic model guidelines \new{state} that regulatory submissions should include detailed descriptions of the models, \remove{including }\new{i.e.~}justification of model assumptions, parameters and their biochemical plausibility, as well as parameter sensitivity analyses~(\citeauthor{EMAreport2011}, \citeyear{EMAreport2011}, p.~16; \citeyear{EMAreport2016}, p.~3).

A key problem is to identify a mathematical model that usefully explains and predicts the behaviour of the pharmacokinetic process \citep{heller2018technologies}. Researchers will often have several models (i.e.\ hypotheses) about an underlying mechanism of the process, but insufficient experimental data to discriminate between the models \couldremove{\citep{heller2018technologies}}. Additional experiments are generally expensive and time-consuming. 
Pharmacokinetic experiments typically take 8--24 hours for mice and rats, and 12--100 hours for simians and humans \citep{Ogungbenro2008, Tuntland2014}. 
\remove{Therefore, i}It is key to design experiments yielding maximally informative outcomes with respect to model discrimination.

\vspace{-2mm}
\paragraph{Contribution}
We tackle limitations in existing analytical and data-driven approaches to design of experiments for model discrimination by bridging the gap between the classical analytical methods and the Monte Carlo-based approaches. These limitations include the classical analytical methods' difficulty to manage non-analytical model functions, and the computational cost for the data-driven approaches. To this end, we develop a novel method of replacing the original parametric models with probabilistic, non-parametric Gaussian process (GP) surrogates learnt from model evaluations. The GP surrogates are flexible regression tools that allow us to extend classical analytical results to non-analytical models in a computationally efficient manner, while providing us with confidence bounds on the model predictions.

\new{The method described in this paper has been implemented in a python software package, \citet{GPdoemd}, publicly available via GitHub under an MIT License.}

\section{Model Discrimination}

Model discrimination aims to discard inaccurate models, i.e., hypotheses that are not supported by the data. Assume ${M>1}$ models $\M_i$ are given, with corresponding model functions $\vec f_i(\vec x, \vec \theta_i)$. The model is a hypothesis and collection of assumptions about a biological system, \new{e.g.~the pharmacokinetic model in the top left of Figure~\ref{fig:pk},} 
and the function is the model's mathematical formulation that can be used for predictions\new{, e.g.~the drug concentrations on the right in Figure~\ref{fig:pk}}. 
Each model function $\vec f_i$ takes as inputs the design variables $\vec x \in \mathcal{X} \subset \R^d$ and model parameters $\vec \theta_i \in \Theta_i \subset \R^{D_i}$. The design variables $\vec x$\new{, e.g.~drug administration and dosing profile,} specify a physical\remove{ (e.g. pharmacokinetic)} experiment that can be run on the system described by the models, with $\mathcal{X}$ defining the operating system boundaries. The classical setting allows tuning of the model parameters $\vec \theta_i$ to fit the model \new{to data}. The system has $E \geq 1$ target dimensions, i.e.\ $\vec f_i:\,\R^{d+D_i} \to \R^E$. We denote the function for each target dimension $e$ as $f_{i,(e)}$. This notation distinguishes between the $e^\mathrm{th}$ target dimension $f_{(e)}$ of a model function $\vec f$, and model $\M_i$'s function $\vec f_i$.
\new{Table~A in the supplementary material summarises the notation used.}

We start with an initial set of $N_0$ experimental measurements $\vec y_n = \vec f_\mathrm{true}(\vec x_n) + \vec \epsilon_n$, $n=1,\dots,N_0$. A common assumption in pharmacokinetics is that the measurement noise term  $\vec \epsilon_n$ is i.i.d. zero-mean Gaussian distributed with covariance $\vec \Sigma_{\mathrm{exp}}$ \citep{Steimer1984, EtteWilliams2004, Tsimring2014}. \new{Skew in the data distribution can be handled, e.g.~through $\log$-transformation.}
The experimental data set is denoted $\Dexp = \lbrace \vec y_j, \vec x_j \rbrace$. The initial $N_0$ experimental data points are insufficient to discriminate between the models, i.e.\ $\forall\,i: p(\M_i\,|\,\Dexp) \approx 1/M$. We are agnostic to the available experimental data, and wish to incorporate all of it in the model discrimination.


\subsection{Classical Analytical Approach}
\label{sec:classicalapproach}
Methods for tackling the design of experiments for discriminating simple, analytical models have been around for over 50 years. \citet{BoxHill1967} study the expected change $\E[\Delta S] = \E[S_{N+1}] - S_N$ in Shannon entropy $S_N = \sum_i \pi_{i,N} \log \pi_{i,N}$ from making an additional experimental observation, where the posterior model probability $\pi_{i,N+1} = p(\vec y|\M_i) \pi_{i,N} / p(\vec y)$ and $p(\vec y) = \sum_i p(\vec y|\M_i) \pi_{i,N}$. \citet{BoxHill1967} develop a design criterion $\DBH$ by maximising the upper bound on $\E[\Delta S]$. The expression for $\DBH$ can be found in the supplementary material. \citet{MacKay1992a, MacKay1992b} derives a similar expression. \citet{BoxHill1967} choose the next experiment by finding $\vec x^\ast = \argmax_{\vec x} \DBH$. Design of experiments continues until there exists a model probability $\pi_{N,i} \approx 1$.

\citet{BuzziFerraris1990} propose a new design criterion $\DBF$ (see the supplementary material) from a frequentist point-of-view. They suggest using a $\chi^2$ test with $NE - D_i$ degrees of freedom for the model discrimination. \new{The null hypothesis for each model is that the model has generated the experimental data.} Design of experiments continues until only one model passes the $\chi^2$ test. Models are not ranked against each other since \citet{BuzziFerraris1990} argue this simply leads to the least inaccurate model being selected. \new{The $\chi^2$ procedure is more robust against---but not immune to---favouring the least inaccurate model.}

\citet{Michalik2010} proceed from the Akaike information criterion (AIC) as the model discrimination criterion to derive a design criterion $\DAW$ from the Akaike weights $w_i$. The expression for $\DAW$ can be found in the supplementary material. Design of experiments continues until there exists an Akaike weight $w_i\approx 1$. \new{\citet{BoxHill1967} and \citet{Michalik2010} implicitly favour the least inaccurate model.}

In order to account for the uncertainty in the model parameters $\vec \theta$, the classical analytical approach \citep{PrasasSomeswaraRao1977, BuzziFerraris1984} is to approximate the model parameters as being Gaussian distributed $\N(\thetamap, \vec \Sigma_\theta)$ around the best-fit parameter values $\thetamap$. The covariance $\vec \Sigma_\theta$ is given by a first-order approximation
{%
\begin{align}
	\label{eq:classicalSigmaTheta}
	\vec \Sigma_\theta\inv
    &= 
    \sum_{n=1}^{N} \nabla_{\vec \theta} {\vec f_i^n}\T \vec \Sigma_{\mathrm{exp}}\inv \nabla_{\vec \theta} \vec f_i^n \,,
\end{align}
}%
where $\vec f_i^n = \vec f_i(\vec x_n, \vec \theta)$, $\vec x_n \in \Dexp$, and the gradient $\nabla_{\vec \theta} \vec f_i = \partial \vec f_i (\vec x, \vec \theta_i) / \partial \vec \theta_i |_{\vec \theta_i=\thetamap_i}$. The predictive distribution with $\vec \theta$ approximately marginalised out becomes $p(\vec f_i \,|\, \vec x, \Dexp)=\N(\vec f_i(\vec x,\thetamap), \Sigmamarg_i(\vec x))$, where the covariance is $\Sigmamarg_i(\vec x) = \nabla_{\vec \theta} \vec f_i \vec \Sigma_\theta \nabla_{\vec \theta} \vec f_i \T$.

\paragraph{Limitations}
Most model functions for healthcare applications such as pharmacokinetics or medical devices are not analytical. They are often, from a practical point-of-view, complex black boxes of legacy code, e.g.\ large systems of ordinary or partial differential equations. We can evaluate the model function $\vec f_i$, but the gradients $\nabla_{\vec \theta} \vec f_i$ are not readily available. Specialist software can retrieve the gradients $\nabla_{\vec \theta} \vec f_i$ from systems of ordinary or partial differential equations, but this requires implementing and maintaining the models in said specialist software packages. Other types of mathematical models may require solving an optimisation problem \new{\citep{BOUKOUVALA2016701}}. For model discrimination, we wish to be agnostic with regards to the software implementation or model type, since this flexibly (i) allows faster model prototyping and development, and (ii) satisfies the personal preferences of researchers and engineers.

\subsection{Bayesian Design of Experiments}
\label{sec:bayesDoE}
Methods for accommodating non-analytical model functions have developed in parallel with increasing computer speed. These methods are typically closer to fully Bayesian than the classical analytical methods. \citet{Vanlier2014} approximate the marginal predictive distributions of $M$ models and their Jensen-Shannon divergence using Markov chain Monte Carlo (MCMC) sampling and a $k$-nearest neighbours density estimate. On the downside, the density estimates become less accurate as the number of experimental observations increases \citep{Vanlier2014} and the method is computationally intensive~\citep{Ryan2016}.

Instead of studying design of experiments for model discrimination, statisticians have focused on design of experiments and model discrimination separately \citep{ChalonerVerdinelli1995}. Design of experiments can also be used to aid model parameter estimation. \remove{Hence, t}\new{T}hey solve problems of the\remove{ general} type
\begin{align}
	\label{eq:BayesDoE}
	\vec x^\ast &= \argmax_{\vec x} \E_{\vec \theta,\, \vec y_{N+1}}
    \left[ U(\vec y_{N+1}, \vec x, \vec \theta) \right]
\end{align}
where \new{the utility function} $U(\cdot,\cdot,\cdot)$ \remove{is a utility function}\new{serves to either discriminate models or estimate parameters}\remove{that can have either a model discrimination or a parameter estimation purpose}. Criteria for model discrimination are handled separately, usually under the name of model selection or hypothesis testing.

\citet{Ryan2015} use a Laplace approximation of the posterior distribution $p(\vec \theta\,|\,\Dexp)$ combined with importance sampling. \citet{Drovandi2015} develop a method based on sequential Monte Carlo (SMC) that is faster than using MCMC. \citet{Woods2017} use a Monte Carlo approximation $\Phi(\vec x) = \frac{1}{B} \sum_{b}^{B} U(\vec y_b, \vec x, \vec \theta_b)$ with $(\vec y_b, \vec \theta_b) \sim p(\vec y_b, \vec \theta_b|\vec x)$ on which they place a Gaussian process prior.

\paragraph{Limitations}
These methods agnostically accommodate non-analytical models using a Monte Carlo-approach but require exhaustive model sampling in the model parameter space. On a case study with four models, each with ten model parameters, and two design variables, the \citet{Vanlier2014} MCMC method requires six days of wall-clock time to compute two sets of four marginal predictive distributions. This cost is impractical for pharmacokinetic applications, where a physical experiment takes \remove{around} $\approx$1--4 days. SMC methods\remove{, e.g.\ \citet{Woods2017},} can converge faster than MCMC methods \new{\citep{Woods2017}}. But \remove{it is well known that }SMC methods can suffer from sample degeneracy\remove{ issues}, where \new{only a few particles receive} the vast majority of the probability weight\remove{ is given to only a few particles} \citep{Li2014}. 
Also, convergence analysis for MCMC and SMC methods is difficult.

Furthermore\remove{, except for \citet{Vanlier2014}}, these methods are only for design of experiments. Once an experiment is executed, the model discrimination issue remains. In this case the marginal predictive distributions $p(\vec f_i\,|\,\vec x, \Dexp)$ would enable calculating the model likelihoods.

\section{Method}
\label{sec:method}
We wish to exploit results from the classical analytical approach to design of experiments for model discrimination, while considering that we may deal with non-analytical model functions. The key idea is replacing the original (non-analytical) model functions with analytical and differentiable surrogate models. The surrogate models can be learnt from model evaluations. We choose to use GPs to construct the surrogate models because they are flexible regression tools that allow us to extract gradient information.

A GP is a collection of random variables, any finite subset of which is jointly Gaussian distributed \citep[p. 13]{RasmussenWilliams2006} and fully specified by a mean function $m(\cdot)$ and a kernel function $k(\cdot, \cdot)$. Prior knowledge about the model functions $\vec f_i$ can be incorporated into the choice of $m$ and $k$. To construct the GP surrogates, target data is needed. The purpose of the surrogate model is to predict the outcome of a model evaluation $f_i(\vec x, \vec \theta_i)$, rather than the outcome of a physical experiment $f_\mathrm{true}(\vec x)$. Hence, for each model $\M_i$ we create a data set $\Dsimi = \lbrace \td{\vec f}_\ell, \td{\vec x}_\ell, \td{\vec \theta}_\ell;\, \ell = 1,\dots,N_{\mathrm{sim},i} \rbrace$ of simulated observations, i.e.\ model evaluations $\td{\vec f}_\ell = \vec f_i (\td{\vec x}_\ell, \td{\vec \theta}_\ell)$, where $\td{\vec x}_\ell$ and $\td{\vec \theta}_\ell$ are sampled from $\mathcal{X}$ and $\Theta_i$, respectively. Note the difference between the experimental data set $\Dexp$ and the simulated data sets $\Dsimi$: whereas the data set $\Dexp$ has few (noisy, biological) observations, $\Dsimi$ has many (noise-free, artificial) observations. 

We place independent GP priors $f_{i,(e)} \sim \GP ( m_{i,(e)},k_{i,(e)} )$ on each target dimension $e=1,\dots,E$ of the model function $\vec f_i$, $i=1,\dots,M$. For notational simplicity, let $f=f_{i,(e)}$, $m=m_{i,(e)}$, $k=k_{i,(e)}$ and $\vec z\T = [\vec x\T, \vec \theta_i\T]$. Given $\Dsimi$ with model evaluations $[\td{\vec f}]_\ell = \td{f}_\ell$ at locations $[\td{\vec Z}]_\ell = \td{\vec z}_\ell$ the predictive distribution $p(f | \vec z_\ast, \td{\vec f}, \td{\vec Z}) = \N(\mu(\vec z_\ast), \sigma^2(\vec z_\ast)$ at a test point $\vec z_\ast$ is given by
{\allowdisplaybreaks
\begin{align}
	\mu (\vec z_\ast) &= m(\vec z_\ast) + \vec k_\ast\T \vec K\inv (\td{\vec f} - \vec m) \,, \\
    \sigma^2 (\vec z_\ast) &= k(\vec z_\ast, \vec z_\ast) - \vec k_\ast\T \vec K\inv \vec k_\ast \,,
\end{align}
}
where the covariance vector is $[\vec k_\ast]_\ell = k(\td{\vec z}_\ell, \vec z_\ast)$, the kernel matrix is $[\vec K]_{\ell_1,\ell_2} = k(\td{\vec z}_{\ell_1},\td{\vec z}_{\ell_2}) + \sigma_n^2 \delta_{\ell_1,\ell_2}$, and the target mean is $[\vec m]_\ell = m(\td{\vec z}_\ell)$, with a small artificial noise variance $\sigma_n^2 \ll 1$ added for numerical stability. Note the simplified notation $\mu = \mu_{i,(e)}$ and $\sigma^2 = \sigma_{i,(e)}^2$.

Pharmacokinetic models are often formulated as systems of ordinary differential equations, e.g.\ \citet{Han2018}, where the solutions are smooth functions. If we wish to model these functions with a GP, it makes sense to use \new{a covariance function that encodes the smoothness of the underlying function, e.g.} the radial basis function (RBF) covariance function 
$k_\mathrm{RBF}(\vec z, \vec z') = \rho \exp \left( -\tfrac{1}{2} (\vec z - \vec z')\T \vec \Lambda\inv (\vec z - \vec z') \right)$, since it corresponds to Bayesian\remove{ generalised} linear regression with an infinite number of basis functions \citep{RasmussenWilliams2006}. \new{Our method extends to other choices of covariance function.}

\new{Each function sample $f$ drawn from a zero-mean GP is a linear combination $f(\vec x) = \sum_j \omega_j k(\vec x, \vec x_j)$ of covariance function terms}~\citep[p. 17]{RasmussenWilliams2006}\new{. \mbox{Different} model characteristics, e.g.~smoothness or periodicity, make alternate covariance functions appropriate. The RBF covariance function is common, but its strong smoothness prior may be too unrealistic for some systems. Other choices include: Mat\'ern, polynomial, periodic, and non-stationary covariance functions. Sums or products of valid covariance functions are still valid covariance functions. \mbox{Covariance} functions can also model different behaviour over different input regions~\citep{AmbrogioniMaris2016}. Choosing the covariance function requires some prior knowledge about the modelled function characteristics, e.g.~a periodic covariance function may be appropriate to model periodic drug dosing.}

\new{In $k_\mathrm{RBF}$, and other similar covariance functions,} $\rho^2$ is the signal variance and $\vec \Lambda = \diag(\lambda^2_{1:D'})$ is a diagonal matrix of squared length scales, with $\vec z \in \R^{D'}$. Together, $\rho$, $\vec \Lambda$ and the artificial noise variance $\sigma_n^2$ (in $\vec K$) are the GP hyperparameters. Because the data set $\Dsimi$ is arbitrarily large, we use point-estimate hyperparameters learnt through standard evidence maximisation.

We study a single model function $\vec f = \vec f_i$ and its GP surrogates. Using the GP surrogates we can predict the outcome of a model evaluation $\vec f (\vec x, \vec \theta) \sim p(\vec f \,|\,\vec z, \Dsimi) = \N(\vec \mu(\vec z), \vec \Sigma(\vec z))$, where $\vec \mu = [\mu_{(1:E)}]\T$ and $\vec \Sigma = \diag(\sigma_{(1:E)}^2)$. We approximate the (marginal) predictive distribution $p(\vec f \,|\, \vec x, \Dsimi, \Dexp)$ to account for the model parameter uncertainty $\vec \theta \sim p(\vec \theta \,|\, \Dexp)$ due to parameter estimation from the noisy experimental data $\Dexp$.

\subsection{Model Parameter Marginalisation}
\label{sec:marginal}
We assume that the experimental observations $\vec y_{1:N} \in \Dexp$ have i.i.d.\ zero-mean Gaussian distributed measurement noise with known covariance $\vec \Sigma_{\mathrm{exp}}$.
Let $\nabla_{\vec \theta} g$ denote the gradient $\partial g (\vec x, \vec \theta) / \partial \vec \theta |_{\vec \theta=\thetamap}$ of a function $g$ with respect to the model parameter $\vec \theta$, evaluated at $\vec \theta=\thetamap$, and $\nabla_{\vec \theta}^2 g$ denote the corresponding Hessian. 

We approximate the model parameter posterior distribution $p(\vec \theta\,|\,\Dexp) \approx \N(\thetamap,\vec \Sigma_\theta)$ as a Gaussian distribution around the best-fit model parameter values $\thetamap$. 
Using the same first-order approximation as the classical analytical method in \eqref{eq:classicalSigmaTheta}, with $\E_{\vec f}[\vec f\,|\,\vec x, \vec \theta] = \vec \mu(\vec x, \vec \theta)$, yields
\begin{align*}
	\vec \Sigma_\theta\inv 
    &=
    \sum_{n=1}^N \nabla_{\vec \theta} {\vec \mu^n}\T \vec \Sigma_\mathrm{exp}\inv \nabla_{\vec \theta} \vec \mu^n \,,
\end{align*}
where $\vec \mu^n = \vec \mu(\vec x_n, \vec \theta)$, with $\vec x_n \in \Dexp$. We approximate the marginal predictive distribution $p(\vec f\,|\,\vec x,\Dsim,\Dexp) \approx \N(\mumarg(\vec x), \Sigmamarg(\vec x))$ using a first- or second-order Taylor expansion around $\vec \theta = \thetamap$, where
\begin{align}
    \label{eq:mumarg}
	\mumarg (\vec x) &= \E_{\vec \theta}
    [\vec \mu\,|\,\vec x,\vec \theta] \\
    \label{eq:Sigmamarg}
    \Sigmamarg (\vec x) &= \E_{\vec \theta} 
    [\vec \Sigma \,|\,\vec x,\vec \theta] 
    + \V_{\vec \theta} 
    [\vec \mu\,|\,\vec x,\vec \theta] \,,
\end{align}
which we will detail in the following.


\paragraph{First-order Taylor approximation} assumes that the model function $\vec f$ is (approximately) linear in the model parameters in some neighbourhood around $\vec \theta = \thetamap$
\begin{align}
    \label{eq:taylorOneMean}
	\mu_{(e)} (\vec x, \vec \theta) 
    &\approx \mu_{(e)} (\vec x, \thetamap) 
    + \nabla_{\vec \theta} \mu_{(e)} (\vec \theta - \thetamap) \,, \\
    \label{eq:taylorOneVariance}
	\sigma_{(e)}^2 (\vec x, \vec \theta)
    &\approx \sigma_{(e)}^2 (\vec x, \thetamap)
    + \nabla_{\vec \theta} \sigma_{(e)}^2 (\vec \theta - \thetamap) \,.
\end{align}
Inserting \eqref{eq:taylorOneMean} into \eqref{eq:mumarg} yields the approximate mean of the marginal predictive distribution
\begin{align}
	\label{eq:taylorOneMumarg}
	\mumarg(\vec x) 
    \approx 
    \vec \mu(\vec x, \thetamap) \,.
\end{align}
Similarly, inserting \eqref{eq:taylorOneVariance} into \eqref{eq:Sigmamarg}, the approximate expectation of the variance becomes
\begin{align}
	\label{eq:taylorOneExpectedVariance}
	\E_{\vec \theta} 
    [\vec \Sigma \,|\,\vec x,\vec \theta] 
    \approx \vec \Sigma(\vec x, \thetamap) \,,
\end{align}
and the variance of the mean becomes
\begin{align}
	\label{eq:taylorOneVarianceMean}
	\V_{\vec \theta} 
    [\vec \mu\,|\,\vec x,\vec \theta]
    \approx
    \nabla_{\vec \theta} \vec \mu 
    \vec \Sigma_\theta
    \nabla_{\vec \theta} \vec \mu \T \,.
\end{align}
\eqref{eq:taylorOneMumarg}--\eqref{eq:taylorOneVarianceMean} yield a tractable first-order Taylor approximation to $p(\vec f\,|\,\vec x, \Dsim, \Dexp) = \N(\mumarg(\vec x),\Sigmamarg(\vec x))$.


\paragraph{Second-order Taylor expansion} assumes that the model function $\vec f$ is (approximately) quadratic in the model parameters in some neighbourhood around $\vec \theta = \thetamap$
\begin{align}
    \begin{split}
    	\mu_{(e)} (\vec x, \vec \theta) 
    	&\approx 
    	\mu_{(e)} (\vec x, \thetamap) 
    	+
    	\nabla_{\vec \theta} \mu_{(e)} (\vec \theta - \thetamap) \\
    	& \quad + 
    	\tfrac{1}{2} (\vec \theta - \thetamap)\T \nabla_{\vec \theta}^2 \mu_{(e)} (\vec \theta - \thetamap) \,,
    	\label{eq:taylorTwoMean}
    \end{split} \\
    \begin{split}
		\sigma_{(e)}^2 (\vec x, \vec \theta) 
    	&\approx 
    	\sigma_{(e)}^2 (\vec x, \thetamap)
    	+
    	\nabla_{\vec \theta} \sigma_{(e)}^2 (\vec \theta - \thetamap) \\
    	& \quad + 
    	\tfrac{1}{2} (\vec \theta - \thetamap)\T \nabla_{\vec \theta}^2 \sigma_{(e)}^2 (\vec \theta - \thetamap) \,.
		\label{eq:taylorTwoVariance}
    \end{split}
\end{align}
Inserting \eqref{eq:taylorTwoMean} into \eqref{eq:mumarg} yields the approximate mean of the marginal predictive distribution
\begin{align}
	\label{eq:taylorTwoMumarg}
	\mumarge(\vec x) 
    &\approx \mu_{(e)}(\vec x, \thetamap) 
    + \tfrac{1}{2} \tr \left( \nabla_{\vec \theta}^2 \mu_{(e)} \vec \Sigma_\theta \right) \,.
\end{align}
Similarly, inserting \eqref{eq:taylorTwoVariance} into \eqref{eq:Sigmamarg} yields the diagonal approximate expectation of the variance with elements
\begin{align}
	\label{eq:expectedVarianceTwo}
    \E_{\vec \theta} 
    [\sigma_{(e)}^2 \,|\,\vec x,\vec \theta]
   	\approx 
    \sigma_{(e)}^2(\vec x, \thetamap) 
   	+
    \tfrac{1}{2} \tr \left( \nabla_{\vec \theta}^2 \sigma_{(e)}^2 \vec \Sigma_\theta \right) \,.
\end{align}
The variance of the mean 
is a dense matrix with elements
\begin{align}
	\begin{split}
		\label{eq:taylorTwoVarianceMean}
		\V_{\vec \theta} 
        [\mu_{(e_1)},\,\mu_{(e_2)}\,|\,&\vec x,\vec \theta]
    	\approx
    	\nabla_{\vec \theta} \mu_{(e_1)}
    	\vec \Sigma_\theta
    	\nabla_{\vec \theta} \mu_{(e_2)}\T \\
        &+
        \frac{1}{2} \tr \left( \nabla_{\vec \theta}^2 \mu_{(e_1)} \vec \Sigma_\theta \nabla_{\vec \theta}^2 \mu_{(e_2)} \vec \Sigma_\theta \right) \,.
	\end{split}
\end{align}
\eqref{eq:taylorTwoMumarg}--\eqref{eq:taylorTwoVarianceMean} yield a tractable second-order Taylor approximation to $p(\vec f\,|\,\vec x, \Dsim, \Dexp) = \N(\mumarg(\vec x),\Sigmamarg(\vec x))$.

\begin{figure*}[!t]
	\vskip 0.2in
	\begin{center}
    	\includegraphics[width=0.95\textwidth]{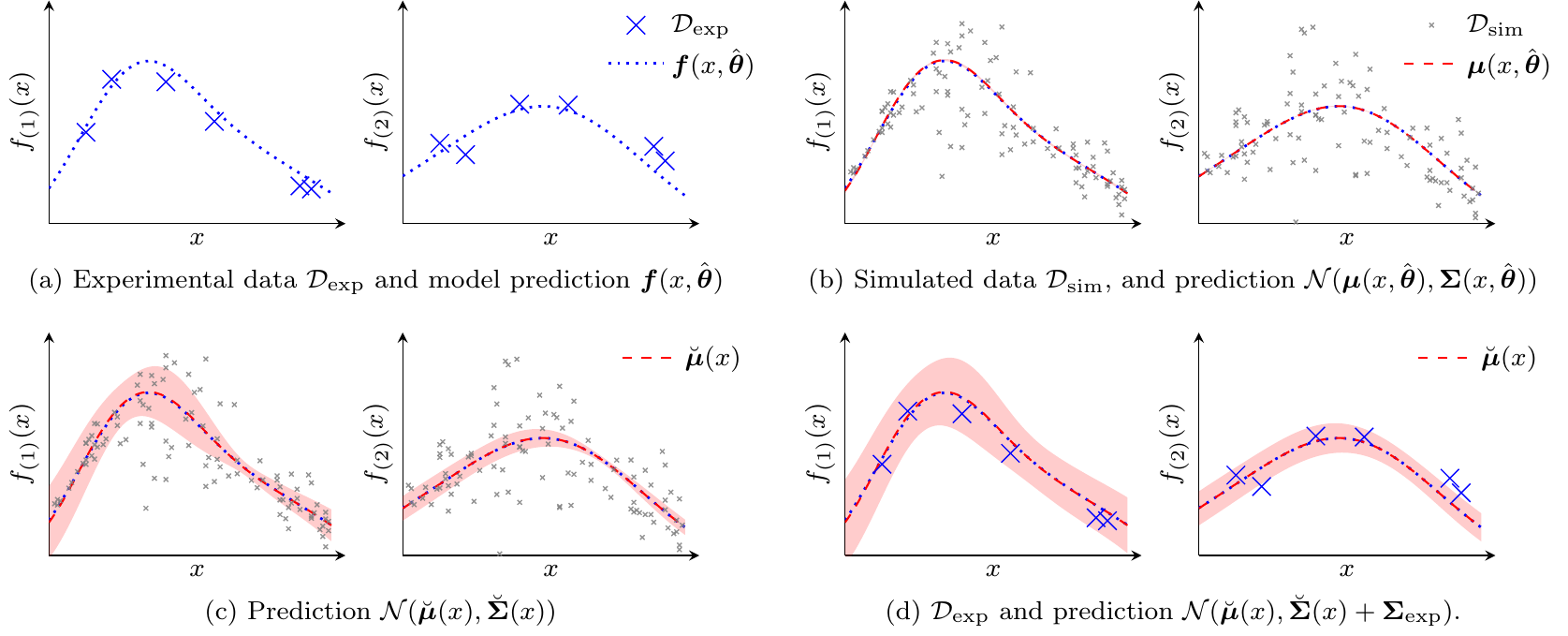}
		\vskip -0.1in
    	\caption{Process of constructing the approximation of the marginal predictive distribution to design a new experiment and perform model discrimination. (a) Step 1, use experimental data $\Dexp$ to perform model parameter estimation and find $\thetamap$. (b) Step 2, generate simulated data $\Dsim$ by sampling from the model $\vec f$. Use $\Dsim$ to learn hyperparameters of the GP surrogate model. (c) Step 3, use Taylor approximation to compute $\mumarg$ and $\Sigmamarg$. Use approximation of marginal predictive distribution to design a new experiment. (d) Step 4, use the approximation of the predictive distribution to compute the model likelihood and perform model discrimination.}
    	\label{fig:GPmarg}
	\end{center}
\end{figure*}

Figure~\ref{fig:GPmarg} illustrates the process of constructing the GP surrogate models and the approximations of the marginal predictive distributions. Firstly, the model parameters are estimated from $\Dexp$. Secondly, $\Dsim$ is created from model evaluations and used (i) to learn the GP surrogate's hyperparameters, and (ii) as target data for GP predictions. Thirdly, the approximation $\N(\mumarg(\vec x),\Sigmamarg(\vec x))$ of the marginal predictive distribution is computed. Fourthly and lastly, the marginal predictive distribution is used to design the next experiment and approximate the model likelihood.

\section{Results}
The following demonstrates that our approach for model discrimination with GP surrogate models exhibits a similar performance to the classical analytical method, but can be used to design experiments to discriminate between non-analytical model functions. The \remove{performance of the }GP surrogate method \new{performance} will be studied using two different case studies. 

Case study 1 has analytical model functions, so we can compare how well our new method performs compared to the classical analytical methods in Section~\ref{sec:classicalapproach}. 
Case study 2 has pharmacokinetic-like models consisting of systems of ordinary differential equations. 


In both case studies we compute the following metrics:%
\setlist[description]{font=\normalfont\space}%
\begin{description}
	\item[(A)] the average number of additional experiments $N-N_0$ required for all incorrect models to be discarded.
	\item[(SE)] the standard error of the average (A).
	\item[(S)] success rate; the proportion of tests in which all incorrect models were discarded.
	\item[(F)] failure rate; the proportion of tests in which the\remove{ the} correct (data-generating) model was discarded.
	\item[(I)] the proportion of inconclusive tests (all models were deemed inaccurate or the maximum number of additional experiments $N_\mathrm{max}-N_0$ was reached).
\end{description}

We compare the design criteria (DC) $\DBH$, $\DBF$ and $\DAW$ described in Section~\ref{sec:classicalapproach} as well as random uniform sampling (denoted Uni.). We also compare the three different criteria for model discrimination (MD) described in Section~\ref{sec:classicalapproach}: updated posterior likelihood $\pi_{N,i}$, the $\chi^2$ adequacy test, and the Akaike information criterion (AIC).

\subsection{Case Study 1: Comparison to Analytical Method}
Case study 1 is from the seminal paper by \citet{BuzziFerraris1984}. With two measured states $y_1$, $y_2$ and two design variables $x_1,x_2 \in [5,55]$, we  discriminate between four chemical kinetic models $\M_i$, each with four model parameters $\theta_{i,j}$ that we fix to $\theta_{i,j} \in [0,1]$:
\begin{alignat*}{2}
	\M_1:\quad f_{(1)} &= \theta_{1,1} x_1 x_2/g_1 \,,\quad
    && f_{(2)} = \theta_{1,2} x_1 x_2/g_1 \,, \\
	\M_2:\quad f_{(1)} &= \theta_{2,1} x_1 x_2/g_2^2 \,,\,
    && f_{(2)} = \theta_{2,2} x_1 x_2/h_{2,1}^2 \,, \\
	\M_3:\quad f_{(1)} &= \theta_{3,1} x_1 x_2/h_{3,1}^2 \,, 
    && f_{(2)} = \theta_{3,2} x_1 x_2/h_{3,2}^2 \,, \\
	\M_4:\quad f_{(1)} &= \theta_{4,1} x_1 x_2/g_4 \,, 
    && f_{(2)} = \theta_{4,2} x_1 x_2/h_{4,1} \,,
\end{alignat*}
where $g_i = 1 + \theta_{i,3} x_1 + \theta_{i,4} x_2$ and $h_{i,j} = 1 + \theta_{i,2+j} x_j$. \citet{BuzziFerraris1984} generate the experimental data $\Dexp$ from model $\M_1$ using $\theta_{1,1}=\theta_{1,3}=0.1$ and $\theta_{1,2}=\theta_{1,4}=0.01$ and Gaussian noise covariance $\vec \Sigma_\mathrm{exp} = \diag(0.35,2.3\mbox{e-}3)$. We start each test with $N_0=5$ randomly sampled experimental observations, and set a maximum budget of $N_\mathrm{max}-N_0=40$ additional experiments. 

Tables~\ref{tab:bf1984grad}--\ref{tab:bf1984t2} compare the performance in 500 completed test instances of the classical analytical approach to the GP surrogate method using first- or second-order Taylor approximations. For this case study, the GP surrogate method performs similarly to the classical analytical approach. We see that the average number of required additional experiments (A) increases using $\pi$ for model discrimination, but that the model discrimination success rate (S) also increases. The model discrimination failure rate drops for almost all MD/DC combinations using the GP surrogate method. We also see that even for this simple case study, the difference in performance between the design criteria and random sampling is significant. The low average (A) for $\pi$/Uni. in Table~\ref{tab:bf1984grad} is likely an artefact due to the large failure rate (F). 
\begin{table}[!b]
	\centering
    \small
	\caption{Results from case study 1 using the analytical approach.}
    \vspace{0.5mm}
    \label{tab:bf1984grad}
\begin{tabular}{c | *{6}{c} }
	\hline
	MD& $\pi$ & $\chi^2$ & AIC & $\pi$ & $\chi^2$ \Tstrut\\
	DC& $\DBH$ & $\DBF$ & $\DAW$ & $\Drandom$ & $\Drandom$ \tstrut\\
	\hline
	A  & 2.60 & 2.87 & 2.08 & 3.43 & 10.39 \Tstrut\\
	SE & 0.04 & 0.12 & 0.04 & 0.11 & 0.45 \tstrut\\
	S [\%] & 86.4 & 64.2 & 62.4 & 59.0 & 85.8 \tstrut\\
	F [\%] & 13.6 & 5.0 & 37.6 & 40.6 & 4.4 \tstrut\\
	I [\%] & 0.0 & 30.8 & 0.0 & 0.4 & 9.8 \tstrut\\
\end{tabular}
    \vspace{2mm}
	\caption{Results from case study 1 using the GP surrogate method with first-order Taylor approximation.}
    \vspace{0.5mm}
    \label{tab:bf1984t1}
\begin{tabular}{c | *{5}{c} }
	\hline
	MD& $\pi$ & $\chi^2$ & AIC & $\pi$ & $\chi^2$ \Tstrut\\
	DC& $\DBH$ & $\DBF$ & $\DAW$ & $\Drandom$ & $\Drandom$ \tstrut\\
	\hline
	A  & 4.31 & 2.23 & 2.72 & 10.99 & 10.02 \Tstrut\\
	SE & 0.09 & 0.06 & 0.08 & 0.34 & 0.45 \tstrut\\
	S [\%] & 95.6 & 47.4 & 88.6 & 69.6 & 83.6 \tstrut\\
	F [\%] & 4.4 & 4.8 & 11.4 & 30.4 & 4.8 \tstrut\\
	I [\%] & 0.0 & 47.8 & 0.0 & 0.0 & 11.6 \tstrut\\
\end{tabular}
    \vspace{2mm}
	\caption{Results from case study 1 using the GP surrogate method with second-order Taylor approximation.}
    \vspace{0.5mm}
    \label{tab:bf1984t2}
\begin{tabular}{c | *{5}{c} }
	\hline
	MD& $\pi$ & $\chi^2$ & AIC & $\pi$ & $\chi^2$ \Tstrut\\
	DC& $\DBH$ & $\DBF$ & $\DAW$ & $\Drandom$ & $\Drandom$ \tstrut\\
	\hline
	A  & 4.14 & 2.29 & 2.64 & 9.53 & 10.11 \Tstrut\\
	SE & 0.09 & 0.07 & 0.06 & 0.30 & 0.47 \tstrut\\
	S [\%] & 96.9 & 46.6 & 90.1 & 64.3 & 80.2 \tstrut\\
	F [\%] & 3.1 & 9.9 & 9.9 & 35.7 & 7.0 \tstrut\\
	I [\%] & 0.0 & 43.5 & 0.0 & 0.0 & 12.8 \tstrut\\
\end{tabular}
\end{table}

\subsection{Case Study 2: Non-Analytical Models}
Case study 2 is from \citet{Vanlier2014}. There are four different models, described by ordinary differential equations. The supplementary material contains the full model descriptions. The models describe a biochemical network, and are similar to typical pharmacokinetic models in (i) the use of reversible reactions, and (ii) the number of parameters and equations (see e.g.\ \citet{Han2018}).

\begin{figure*}[!ht]
	\vskip 0.2in
	\begin{center}
        \includegraphics[width=\textwidth]{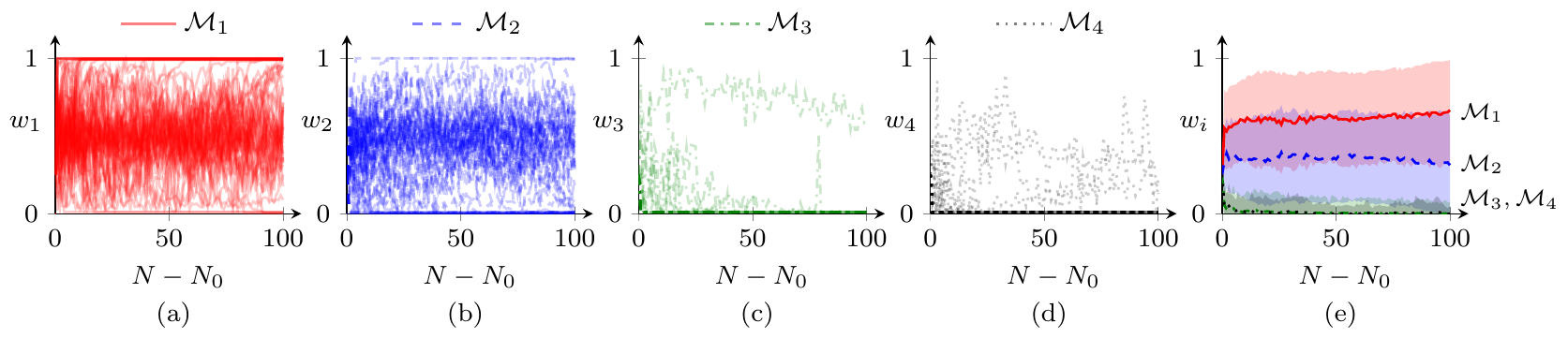}
		\vskip -0.11in
    	\caption{Results from case study 2 with 4 models using the $\DAW$ design criterion and AIC for model discrimination. (a)--(d) show the evolution of the Akaike weights $w_i$ for model $\M_1,\dots,\M_4$, respectively, for 63 completed tests. (e) shows the average Akaike weight evolutions (with one standard deviation) in (a)--(d). The results indicate that models $\M_1$ and $\M_2$ are indiscriminable.}
    	\label{fig:vanlier}
	\end{center}
\end{figure*}

We assume two measured states ($E=2$), with model parameters $\vec \theta_i \in [0,1]^{10}$, initial concentrations $\vec c \in [0,1]^E$, measurement time point $t \in [0,20]$, and design variables $\vec x\T = [t, \vec c\T]$. We follow \citet{Vanlier2014} and let $\M_1$ generate the observed experimental data, with random uniformly sampled true model parameters and experimental measurement noise covariance $\vec \Sigma_\mathrm{exp} = 9.0\times10^{-4}\vec I$. Each test is initiated with $N_0 = 20$ observations at random design locations $\vec x_1, \dots, \vec x_{N_0}$. We set a maximum budget of $N_\mathrm{max}-N_0=100$ additional experiments.

Case study 1 shows no obvious benefit in using the second-order over a first-order Taylor approximation. Hence, in case study 2 we only performance test the GP surrogate method with the first-order Taylor approximation.

The results in Table~\ref{tab:vanlier} for 4 models show that the rates of inconclusive results are higher for case study 2 than for case study 1.
\begin{table}[!b]
	\centering
    \small
    \caption{Results from case study 2.}
    \vspace{0.5mm}
    \label{tab:vanlier}
\begin{tabular}{c | *{3}{c} | *{3}{c} }
        \hline
        & \multicolumn{3}{c|}{\emph{4 models}} & \multicolumn{3}{c}{\emph{3 models}} \Tstrut\\
        MD& $\pi$ & $\chi^2$ & AIC & $\pi$ & $\chi^2$ & AIC \tstrut\\
        DC& $\DBH$ & $\DBF$ & $\DAW$ & $\DBH$ & $\DBF$ & $\DAW$ \tstrut\\
        \hline
        A  & 20.10 & 39.83 & 29.62 & 15.80 & 21.91 & 9.74 \Tstrut\\
        SE & 3.72 & 12.09 & 7.72 & 2.05 & 2.52 & 1.70 \tstrut\\
        S [\%] & 15.9 & 9.5 & 33.3 & 89.5 & 77.2 & 95.6 \tstrut\\
        F [\%] & 7.9 & 0.0 & 7.9 & 6.1 & 0.9 & 1.8 \tstrut\\
        I [\%] & 76.2 & 90.5 & 58.7 & 4.4 & 21.9 & 2.6 \tstrut\\
\end{tabular}
\end{table}
Furthermore, the ratio of successful model discriminations to failed model discriminations are lower for $\pi$/$\DBH$ and AIC/$\DAW$ than they are in case study 1. This is because for case study 2 the experimental noise is often larger than the difference between the model predictions. Figure~\ref{fig:vanlier} shows the evolutions of the Akaike weights $w_i$ \citep{Michalik2010} for all 63 completed tests.
The Akaike weight evolutions show that especially models $\M_1$ and $\M_2$ are difficult to discriminate between, given that we only consider two measured states. Corresponding graphs for $\pi$ and $\chi^2$ are similar-looking. One can also see from the model formulations (see supplementary material) that all models are similar.

To verify that the Table~\ref{tab:vanlier} results for 4 models are indeed due to the indiscriminability of model $\M_1$ and $\M_2$, we carry out another set of tests where we only consider models $\M_1$, $\M_3$ and $\M_4$. The results in Table~\ref{tab:vanlier} for 3 models show that removing model $\M_2$ enables the GP surrogate method to correctly discriminate between the remaining models more successfully. In practice, researchers faced with apparent model indiscriminability have to rethink their experimental set-up to lower measurement noise or add measured states.

\section{Discussion}
The healthcare industry already use\new{s} machine learning to help discover new drugs \new{\citep{Pfizer2016}}. 
But passing the pre-clinical stage is expensive. \citet{DiMasi2016} estimate the average pre-approval R\&D cost to \$2.56B (in 2013 U.S. dollars) with a yearly increase of 8.5\,\%. For the pre-clinical stage (before human testing starts), \citet{DiMasi2016} estimate the R\&D cost to \$1.1B. 

Model discrimination, which takes place in the pre-clinical stage, may lower drug development cost by identifying a model structure quickly. \citet{heller2018technologies} explain that a major hurdle for \emph{in silico} pharmacokinetics is the difficulty of model discrimination. \citet{ScannelBosley2016} and \citet{Plenge2016} also attribute a significant part of the total R\&D cost for new drugs to inaccurate models passing the pre-clinical stage only to be discarded in later clinical stages. Inaccurate models need to be discarded sooner rather than later to avoid expensive failures.

Design of experiments for model discrimination has typically focussed either on analytical approaches, which cannot manage all functions, or on data-driven approaches, which may have computational difficulties or lack interpretable marginal predictive distributions. We leverage advantages of both approaches by hybridising the analytical and data-driven approaches, i.e.\ we replace the analytical functions from the classical methods with GP surrogates.

The novel GP surrogate method is highly effective.
Case study 1 directly compares \new{it} with the analytical approach on a classical test instance. The similar performance indicates successful hybridisation between analytical and data-driven approaches.
Case study 2 considers an industrially relevant problem suffering from model indiscriminability. The GP surrogate method successfully avoids introducing overconfidence into the model discrimination despite the assumptions used to derive the marginal predictive distribution. 

The parameter estimation step is limited to few and noisy data points, e.g.\ from time-consuming pharmacokinetic experiments. If the approximation $p(\vec \theta\,|\,\Dexp) \approx \N(\thetamap, \vec \Sigma_\theta)$ is not accurate, the Section~\ref{sec:marginal} Taylor approximations of $\mumarg$ and $\Sigmamarg$ can be extended to a mixture of Gaussian distributions $p(\vec \theta\,|\,\Dexp) = \sum_j \omega_j \N(\thetamap_j, \vec \Sigma_{\theta,j})$, with $\sum_j \omega_j = 1$, if $\lbrace \thetamap_j, \vec \Sigma_{\theta,j} \rbrace$ are given. \new{The analytical approximations sacrifice some accuracy for computational tractability:} Our computational time for design of experiments in case study 2 was on the order of minutes, whereas the method of \citet{Vanlier2014} requires days. \new{Our results indicate that the approximations work well in practice.}
GP surrogate methods could sharply reduce the cost of drug development's pre-clinical stage, e.g.\ by shrinking the time needed to decide on a new experiment or by finding a good model structure in a reduced number of pharmacokinetic experiments.

Our new method directly applies to neighbouring fields. For example, dual control theory deals with control of unknown systems, where the goal is to balance controlling the system optimally vs.\ learning more about the system. When rival models predict the unknown system behaviour, learning about the system involves designing new experiments, i.e.\ control inputs to the system \citep{CheongManchester2014}. This design of experiments problem, e.g.\ in \citet{larsson2013model}, would be likely to have very strict time requirements and benefit from confidence bounds on the model predictions. \new{Our method also applies to neighbouring application areas, e.g.~distinguishing between microalgae metabolism models \citep{ULMASOV20161051} or models of bone neotissue growth \citep{mehrian-etal:2018}.}

A very hot topic in machine learning research is learning in implicit generative models \citep{Shakir2017}, e.g.\ generative adversarial networks (GANs) \citep{Goodfellow2014}, which employ a form of model discrimination. Here the likelihood function $p(\vec y\,|\,\M_i)$ is not given explicitly, so the goal is to learn an approximation $q(\vec y\,|\,\M_i)$ of the data distribution $p(\vec y)$. The model discrimination tries to identify from which distribution $q(\vec y\,|\,\M_i)$ or $p(\vec y)$ a random sample has been drawn. For the case of design of experiments for model discrimination, we have a prescribed probabilistic model, since we explicitly specify an approximation of the likelihood $p(\vec y\,|\, \M_i, \vec x, \vec \theta_i)$. But we have (i) the physical experiments, the true but noisy data generator that is expensive to sample, and (ii) the rival models, deterministic and relatively cheap (but probably inaccurate!) data generators. Design of experiments for model discrimination also ties in with several other well-known problems in machine learning, e.g.\ density estimation, distribution divergence estimation, and model likelihood estimation.

\section{Conclusions}
We have presented a novel method of performing design of experiments for model discrimination. The method hybridises analytical and data-driven methods in order to exploit classical results while remaining agnostic to the model contents. Our method of approximating the marginal predictive distribution performs similarly to classical methods in a case study with simple analytical models. For an industrially-relevant system of ordinary differential equations, we show that our method both successfully discriminates models and avoids overconfidence in the face of model indiscriminability. The trade-off between computational complexity and accuracy that we introduce is useful for design of experiments for discrimination e.g.\ between pharmacokinetic models, where the duration of the experiments is shorter than or one the same time-scale as the computational time for Monte Carlo-based methods.

\section*{Acknowledgements} 
This work has received funding from the European Union's Horizon 2020 research and innovation programme under the Marie Sk\l{}odowska-Curie grant agreement no.675251, and an EPSRC Research Fellowship to R.M. (EP/P016871/1).

\bibliography{ref}
\bibliographystyle{icml2018}

\clearpage
\pagebreak
\appendix

\section{\new{Table of Notation}}
\setcounter{table}{0}
\renewcommand{\thetable}{\Alph{table}} 
\begin{table}[!h]
	\centering
    \vspace{-2mm}
    \caption{Summary of notation.}
    \label{tab:notation}
    \begin{tabular}{c l}
    \hline
    Symbol & Description \Tstrut\Bstrut\\ 
    \hline
    $\vec f_i$ & Function associated with model $\M_i$
    \Tstrut\Bstrut\tablespace
    $f_{i,(e)}$ & Dimension $e$ of function $\vec f_i$; $e=1,\dots,E$
    \tstrut\Bstrut\tablespace
    $\vec x$ & Design variable, $\vec x \in \mathcal{X} \subset \R^d$
    \tstrut\Bstrut\tablespace
    $\vec \theta_i$ & Parameters of model $\M_i$ , $\vec \theta_i \in \Theta_i \subset \R^{D_i}$
    \tstrut\Bstrut\tablespace
    $M$ & No. of models $\M_i$;\, $i = 1, \dots, M$
    \tstrut\Bstrut\tablespace
    $E$ & No. of target dimensions; $\vec f_i:\,\R^{d+D_i} \to \R^E$
    \tstrut\Bstrut\tablespace
    $\vec \Sigma_\mathrm{exp}$ & Measurement noise covariance
    \tstrut\Bstrut\tablespace 
    $\Dexp$ & The set of experimental observations
    \tstrut\Bstrut\tablespace
    $\Dsimi$ & The set of simulated data for model $\M_i$
    \tstrut\Bstrut\tablespace 
    
    \hline
    \end{tabular}
\end{table}

\section{Design Criteria}

We let $\Delta_{ij} = \vec f_i(\vec x, \thetamap_i) - \vec f_j(\vec x, \thetamap_j)$ and the covariance $\vec \Sigma_i = \vec \Sigma_{\mathrm{exp}} + \Sigmamarg_i(\vec x)$, where $\Sigmamarg_i(\vec x)$ is the covariance of model $\M_i$'s marginal predictive distribution due to model parameter uncertainty.

For a single-response system, \citet{BoxHill1967} derive the design criterion $\DBH$, later generalised to a multi-response form by \citet{PrasasSomeswaraRao1977}
\begin{align*}
	\begin{split}
        \DBH(\vec x) 
        = 
        \sum_{i,j=1}^{M} & \frac{\pi_{N,i} \pi_{N,j}}{2} \left\lbrace \Delta_{ij}\T \left( \vec \Sigma_i\inv + \vec \Sigma_j\inv \right) \Delta_{ij} \right. \\
        &+
        \left. \tr \left( \vec \Sigma_i \vec \Sigma_j\inv + \vec \Sigma_j \vec \Sigma_i\inv - 2 \vec I \right) \right\rbrace \,.
	\end{split}
\end{align*}

\citet{BuzziFerraris1990} derive the design criterion $\DBF$
\begin{align*}
	\begin{split}
		\DBF (\vec x) 
        = 
        \max_{1 \leq i,j \leq M} &\left\lbrace \Delta_{ij}\T \left( \vec \Sigma_i + \vec \Sigma_j  \right)\inv \Delta_{ij}  \right. \\
        &+
        \left. \tr \left( 2 \vec \Sigma_{\mathrm{exp}} \left( \vec \Sigma_i + \vec \Sigma_j  \right)\inv  \right) \right\rbrace \,.
	\end{split}
\end{align*}
designed such that if $\max_{\vec x} \DBF (\vec x) < E$, the largest difference between model predictions is too small compared to the measurement noise variance to carry out model discrimination, and design of experiments terminates.

\citet{Michalik2010} proceed from the Akaike information criterion (AIC) as the model discrimination criterion to derive a design criterion $\DAW$ from the Akaike weights
\begin{align*}
	w_i(\vec x) &= \frac{1}{ \sum_{j=1}^M \exp \left( \tfrac{-1}{2} \Delta_{ij}\T \vec \Sigma_{i}\inv \Delta_{ij} + D_i - D_j \right) } \,,
\end{align*}
yielding $\DAW = \sum_i w_i p(\M_i)$, where $p(\M_i)$ is the prior probability of model $\M_i$.

\section{Case study from \citet{Vanlier2014}}
There are nine chemical components with concentrations $C_i$, $i=1,\dots,9$. The system of ordinary differential equations has the form
\begin{align*}
	\begin{array}{rrrrlrrrl}
        \mathrm{d} C_1 / \mathrm{d} t & = & -g_1 & + & g_2 \,, \\
        \mathrm{d} C_2 / \mathrm{d} t & = &  g_1 & - & g_2 \,, \\
        \mathrm{d} C_3 / \mathrm{d} t & = & -g_3 & + & g_4 \,, \\
        \mathrm{d} C_4 / \mathrm{d} t & = &  g_3 & - & g_4 & - & g_5 & + & g_6 \,, \\
        \mathrm{d} C_5 / \mathrm{d} t & = & -g_9 & + & g_{10} \,, \\
		\mathrm{d} C_6 / \mathrm{d} t & = & -g_5 & + & g_6 & + & g_9 & - & g_{10} \,, \\
        \mathrm{d} C_7 / \mathrm{d} t & = &  g_5 & - & g_6 \,, \\
        \mathrm{d} C_8 / \mathrm{d} t & = & -g_7 & + & g_8 \,, \\
        \mathrm{d} C_9 / \mathrm{d} t & = &  g_7 & - & g_8 \,,
	\end{array}
\end{align*}
i.e.\ the stoichiometry is the same for all models $\M_i$. But some of the fluxes $g_1,\dots,g_{10}$ differ for the different models. For all models $\M_i$ the following fluxes are identical:
\begin{align*}
	\begin{array}{lcl r lcl}
		g_2 & = & \theta_{i,2} C_2 \,,		& \hspace{5mm} & g_7 & = & \theta_{i,7} C_8 \,, \\
        g_4 & = & \theta_{i,4} C_4 \,,		& & g_8 & = & \theta_{i,8} C_9 \,, \\
        g_5 & = & \theta_{i,5} C_4 C_6 \,,	& & g_9 & = & \theta_{i,10} C_5 \,, \\
        g_6 & = & \theta_{i,6} C_7  \,,		& & g_{10} & = & \theta_{i,4} C_6 \,.
	\end{array}
\end{align*}
For flux $g_1$ the models differ in the following way:
\begin{align*}
	\M_i: \quad &g_1 = \theta_{i,1} C_1 \,, \quad i \in \lbrace 1,3,4 \rbrace \\
	\M_2: \quad &g_1 = \frac{\theta_{2,1} C_1}{\theta_{2,9} + C_7} \,.
\end{align*}
For flux $g_3$ the models differ in the following way:
\begin{align*}
	\begin{array}{lrl}
		\M_1:\,\, g_3 = \frac{\theta_{1,3} C_2 C_3}{\theta_{1,9} + C_7} \,, & & 
        \M_3:\,\, g_3 = \frac{\theta_{3,3} C_2 C_3}{\theta_{3,9} + C_9} \,, \\
        \M_2:\,\, g_3 = \theta_{2,3} C_2 C_3 \,, & & 
        \M_4:\,\, g_3 = \frac{\theta_{4,3} C_2 C_3}{\theta_{4,9} + C_8} \,.
	\end{array}
\end{align*}

We assume that the only measured states are the concentrations $C_4$ and $C_9$, because these are the states from which \citet{Vanlier2014} collect their initial data. Similarly, we use the initial concentrations $C_4(t=0)$ and $C_9(t=0)$ as two of our design variables, the third design variable being the time point $t$ at which to measure the concentrations. 

\citet{Vanlier2014} look at times points in the range $t \in [0,20]$, which we also adopt. We assume the initial concentrations $C_4(t=0),C_9(t=0) \in [0,1]$ and fix all other initial concentrations to
\begin{align*}
	C_1(t=0) = C_3(t=0) = C_5(t=0) = C_8(t=0) &= 1 \,, \\
	C_2(t=0) = C_6(t=0) = C_7(t=0) &= 0.1 \,.
\end{align*}
We assume the model parameter space $\vec \theta \in [0,1]^{10}$. Simulations show that sampling from this parameter space gives a wide range of model realisations.

With reference to models $\M_1$ and $\M_2$ being similar, we see that the only difference between them is that the term $\theta_{i,9} + C_7$ divides $g_1$ and $g_3$ for $\M_1$ and $\M_2$, respectively. If $C_7$ is small compared to $\theta_{i,9}$, then the models are nearly identical.

\end{document}